\documentclass[10pt, conference, compsocconf]{IEEEtran}

\usepackage[utf8]{inputenc} 
\usepackage[T1]{fontenc}    
\usepackage{url}            
\usepackage{booktabs}       
\usepackage{amsfonts}       
\usepackage{nicefrac}       
\usepackage{microtype}      
\usepackage{lipsum}
\usepackage{graphicx}
\usepackage{amsmath}
\usepackage{tikz}
\usepackage{listings}
\usetikzlibrary{automata}
\graphicspath{ {./images/} }

\makeatletter
\def\BState{\State\hskip-\ALG@thistlm}
\makeatother

\definecolor{codegreen}{rgb}{0,0.6,0}
\definecolor{codegray}{rgb}{0.5,0.5,0.5}
\definecolor{codepurple}{rgb}{0.58,0,0.82}
\definecolor{backcolour}{rgb}{0.95,0.95,0.92}

\begin{document}

\title{Security Vulnerability Detection Using Deep Learning Natural Language Processing}

\author{
\IEEEauthorblockN{Noah Ziems} 
\IEEEauthorblockA{
Department of Computer Science \\
Ball State University\\
nmziems@bsu.edu
}
\and
\IEEEauthorblockN{Shaoen Wu}
\IEEEauthorblockA{
School of Information Technology \\
Illinois State University \\
swu1235@ilstu.edu}

}

\maketitle
\begin{abstract}
Detecting security vulnerabilities in software before they are exploited has been a challenging problem for decades. Traditional code analysis methods have been proposed, but are often ineffective and inefficient. In this work, we model software vulnerability detection as a natural language processing (NLP) problem with source code treated as texts, and address the automated software venerability detection with recent advanced deep learning NLP models assisted by transfer learning on written English. For training and testing, we have preprocessed the NIST NVD/SARD databases and built a dataset of over 100,000 files in $C$ programming language with 123 types of vulnerabilities. The extensive experiments generate the best performance of over 93\% accuracy in detecting security vulnerabilities.
\end{abstract}

\section{Introduction}\label{introduction}

Some recent advancements, called transformers \cite{vaswani2017attention,kaiser2017model,kaiser2017depthwise}, in Natural Language Processing (NLP) show great promise in the ability to extract situational meaning from structured text. The transformers employ a sequence-to-sequence architecture where each layer simply outputs a \textit{transformation} of the sequence it is given. Traditional models instead rely on shallow \textit{memorization} mechanisms which can be prone to forgetting important information, particularly with longer sequences. By simply doing a sequence-to-sequence transformation at each layer, transformers are able to have far more layers than previous state-of-the-art sequence models. In turn, this deeply-stacked transformation architecture allows transformers to outperform traditional models on a wide variety of tasks. The numerous applications of these transformers range from text classification to question answering. Moreover, they have been proven to be agnostic with respect to language-structure, performing well not only on English text but also other languages such as Arabic, Chinese(Traditional), and others.

Although these models have been extensively applied to and evaluated in the context of human languages, there is very little work of transformers for other text applications. As far as we know, there has been no report of the success of these transformer-based models on non-language text applications. This work takes the effort of designing high-performing transformer models with raw computer code as the input sequence.

While modern programming languages are able to detect syntax and runtime errors in real-time development environments, detecting security errors is quite challenging during software development. This is likely due to the large number of different security vulnerabilities and the complicated conditions that each one must satisfy to qualify as vulnerable. In worst cases, security vulnerabilities are detected only after they have been exploited, often months or years after the original code was developed. However, common security vulnerabilities often have defined structures and patterns, which suggests that they can be detected with NLP in real-time as they are being developed rather than later, after they have been exploited by bad actors.

While there are a variety of traditional techniques used to detect errors and other issues within code \cite{mahmood2018evaluation, yang2020recognize}, these techniques are plagued with problems, often being inaccurate or extremely misleading. Static code analysis looks at the raw code for hard-coded patterns that indicate security vulnerabilities. Dynamic code analysis runs the raw code in an effort to find problems with the underlying computation behavior. In this work, we employ deep machine learning to extract security patterns in code that lend themselves to errors and other security issues. Previous work has been done on this problem with some success \cite{russell2018automated}, \cite{mokhov2014}, but as far as we know we are the first to approach this problem using state-of-the-art transformer-based architectures.

Our main contributions are as follows:
\begin{enumerate}
    \item First, we have built a large open text-based dataset comprised of raw C/C++ code.
    \item Second, we have developed and tested deep learning models to be effective at extracting contextual information from raw code. In particular, we have shown the ability of Bidirectional Encoder Representations from Transformers(BERT)\cite{devlin2018bert} to classify security vulnerabilities in raw code written in C/C++, vastly outperforming the other models tested. 
    \item Third, we have successfully applied transfer learning from written English to computer code and observed that it is surprisingly effective despite their respective structures being significantly different.
\end{enumerate}

The rest of this paper is organized as follows. In Section \ref{relatedwork} we give an overview of the recent work in software vulnerability detection using both static and dynamic code analysis as well as the related deep learning work in transformers and using machine learning to extract features from computer code. Section \ref{dataset} presents the details of the open C code dataset we have built, its raw sources, and how it was created. In Section \ref{models}, we describe in detail the deep learning models we used for software vulnerability detection. Then, Section \ref{performance} compares the performance of these modules on software vulnerability detection with the dataset we developed.

\section{Related Work} \label{relatedwork}
The works that relate to ours can be split up into two main categories: the NLP deep learning models for human languages, and the traditional software vulnerability detection solutions. 

\subsection{Deep Learning in NLP}
The last decade has witnessed great progress using deep learning in NLP applications, such as Google Translate V3 and smart voice assistance. Recently, a new NLP architecture family emerged called {\it transformers} which are much more efficient at computing the complex relationships seen in natural language. Moreover, transformers are able to borrow strategies previously seen in the field of Computer Vision(CV) to more effectively and efficiently train for downstream tasks.

In CV deep learning, the ImageNet challenge \cite{russakovsky2014imagenet} asks its participants to create models that take in a 224 by 224 image with 3 color channels and classify the main subject, also known as the \textit{class} of the image. In total there are over 20,000 classes covering a broad spectrum of objects seen in day-to-day life. While there are now models which achieve impressive performance on this task, they often take more than a week of computation time on a cluster of GPUs to achieve these results\cite{xie2016aggregated}. CV practitioners and researchers can then fine-tune these pre-trained models on new tasks or datasets with relatively little computation time, often achieving surprising results. This process is known as \textit{transfer learning} \cite{razavian2014cnn}.

Until recently, relatively little research had been done on using the ideas behind transfer learning and applying them to the field of NLP. However, new methods using a process of \textit{pre-training} and \textit{fine-tuning} \cite{howard2018universal}, which share similar mechanisms to transfer learning in CV, have shown to produce impressive results on a wide variety of tasks from classification to semantic analysis. During the pre-training phase, an NLP model is asked to predict a masked word in a sequence in an unsupervised manner. In doing so, the model learns the underlying structure of the text it is working with before transitioning to the fine-tuning phase, where it is asked to do the intended task. In NLP, the pre-training phase is akin to training a model on ImageNet for days while the fine-tuning phase is akin to training the model on a new task or dataset.

New transformer-based deep learning models coupled with pre-training and fine-tuning have been shown to be extremely effective at a variety of NLP tasks while also being far easier to train. Bidirectional Encoder Representations from Transformers(BERT)\cite{devlin2018bert} achieves state-of-the-art performance on eleven different NLP tasks including acceptability, semantic analysis, and sentence similarity\cite{wang2018glue} by looking at a sequence in a bidirectional context instead of a unidirectional context. In doing so, this allows words later in a sequence to change the meaning of the previous words and vice versa.

Further, a great deal of work has been done exploring the language-agnostic properties of BERT including the ability to perform well on Arabic \cite{antoun2020arabert} and Chinese \cite{cui2019pretraining} texts. Other transformer models with similar attention-based architectures show further improvements. XLNet uses an auto-regressive pre-training method that learns bidirectional contexts, outperforming BERT on 20 separate NLP tasks \cite{yang2019xlnet}. Transformer-XL \cite{dai2019transformerxl} solves the fixed length context problems shown by other transformer architectures, allowing inputs of long sequences without loss in performance. Transformer-XL was shown to outperform vanilla transformer models on both short and long sequences while being three orders of magnitude faster during evaluation.

Moreover, there has been previous work investigating the ability of NLP models to extract valuable features from computer code. Karpathy et al. \cite{karpathy2015visualizing} have trained an Recurrent Neural Network (RNN) to predict the next character in a given sequence of {\it C} code, then analyze the hidden activation. In doing so, they found the RNN was able to learn a variety of important structures about the code including if-statements, loops, and nesting. Others have used more traditional NLP models in an attempt to detect security vulnerabilities in software code \cite{russell2018automated}, \cite{mokhov2014} showing the potential of NLP to detect security vulnerabilities.

\subsection{Security Vulnerability Detection}
There are a variety of traditional ways in which security vulnerabilities can be detected. However, for the most part they fall into one of two categories. One traditional technique called \textit{Static Code Analysis} analyses the source code of a program looking for places where it breaks any hard-coded rules \cite{Wichmann1995IndustrialPO}. For example, it looks for any unused or duplicated code that may need to be fixed. However, these hard-coded static code analysis tools tend to be error-prone. Often times, when evaluating code for security vulnerabilities, they give many more false-positives than true-positives, causing programmers to chase down a variety of false leads or give up on detecting those actual vulnerabilities all together. Another traditional technique known as \textit{Dynamic Code Analysis} attempts to find issues in the code by running it on the processor with a variety of extreme values as inputs then looking for inconsistencies or errors in the outputs. However, these tests often need to be written by the programmer in the form of unit tests and integration tests. While dynamic code analysis is less prone to false-positives than static code analysis, it is cumbersome to use and often tests inputs beyond the range of what can be reasonably expected of the code. Moreover, dynamic code analysis can only test issues that the programmer has anticipated, which is often not the case for security vulnerabilities.

\section{Development of Software Venerability Dataset} \label{dataset}
Because little work has been done using NLP to detect software vulnerabilities in code, there are few existing datasets for us to reuse. Thus, we build a custom dataset of security vulnerabilities to train and test our models. While there are a variety of different datasets to start from, we have chosen to use the Software Assurance Reference Dataset (SARD) created by the National Institute of Standards and Technology (NIST) as our base. SARD is particularly attractive because it contains not only the security vulnerabilities, but also their non-vulnerable counterparts, allowing our models to learn exactly what the difference is between a security vulnerability and its safe alternatives. We then preprocess the dataset before training by removing any unwanted artifacts that may cause our models to overfit. We end up with over 100,000 files, with half containing one of over 100 security vulnerabilities and the other half contain their non-vulnerable alternatives.

\subsection{Software Assurance Reference Dataset (SARD)}
The SARD database is made up of text files containing raw code. Each file is labeled with a particular Common Weakness Enumeration (CWE) tag corresponding to the security vulnerability that the code contains. While there are a wide variety of languages in the SARD database, we only test and train our models on the subset of code written in C/C++. For each file that contains a vulnerability labeled by CWE, there is a matching file containing code that has been fixed and no longer has the vulnerability. From this, half of our dataset is made up of files without security vulnerabilities and the other half is made up of matching files with vulnerabilities. In total, there are 123 types (lables) of CWE in the dataset. After adding one more label to classify files as non-vulnerable, our dataset has a total of 124 unique labels.

\subsection{Preprocessing}
In the raw SARD database, the specific functions containing security vulnerabilities have the word "bad" appended to the end of the function and variable names, as shown below. The files which contain the non-vulnerable alternatives have the word "good" appended in the same manner. If left uncleaned, our models would simply learn to look for the word "good" or "bad" in the file in order to classify it properly. To fix this, we simply remove all instances of the word "good" or "bad" from all of the files using a very simple obfuscation technique. In an effort to remove any unnecessary noise from the data, we remove all comments as well. Lastly, we format each file in the same manner to maintain a consistent code structure. In practice, all of this should be done during preprocessing before inference time.

\begin{lstlisting}[language=C, caption=SARD sample before preprocessing]
void CWE121_Buffer_Overflow_badSink(void * data)
{
    {
        /**
            POTENTIAL FLAW:
            treating pointer as a char* when it may point to a wide string
        */
        size_t dataLen = strlen((char *)data);
        void * dest = (void *)ALLOCA((dataLen+1) * sizeof(wchar_t));
        (void)wcscpy(dest, data);
        printLine((char *)dest);
    }
}
\end{lstlisting}

\begin{lstlisting}[language=C, caption=Listing 1 after preprocessing]
void func1(void * data)
{
    size_t dataLen = strlen((char *)data);
    void * dest = (void *)ALLOCA((dataLen+1) * sizeof(wchar_t));
    (void)wcscpy(dest, data);
    printLine((char *)dest);
}
\end{lstlisting}

\section{Deep Learning Software Venerability Detection} \label{models}
In this work, we have developed a few deep learning models based on classic NLP architectures for software vulnerability auto detection, including {\it an LSTM model, a bi-directional LSTM model,} and {\it a transformer BERT model}, in an effort to sufficiently and fairly compare latest models to other techniques. Each model's objective is to classify a file of code. It is worth noting that this is not an exhaustive list of models that could be used.

\subsection{Inputs and Outputs of Models}\label{subsec:inout}
The inputs to our our deep learning software vulnerability models are taken in as a long character string comprising of an entire file of code written in C. The string is then broken into words and sub-words by a tokenizer. Note that syntax characters like parentheses, brackets, periods, and semicolons are considered separate words. The words are then coded into representation vectors by the encoder. Depending on which model is being used, these word vector representations are then fed into the model token by token or in large chunks of tokens.

In our deep learning software vulnerability models, the output vector has the same number of dimensions as the number of vulnerability classes in the dataset. Namely, since there are 124 different classes of vulnerability, each output is a vector with 124 dimensions. A Softmax function is then applied to normalize the output vector into a probability distribution where all the probabilities sum to ``1" and each value is the predicted probability that the given code file contains that corresponding vulnerability.

\subsection{Long Short Term Memory (LSTM) Model} \label{lstm}
Long Short Term Memory (LSTM)\cite{HochSchm97} models use a different recurrence formula than vanilla RNNs that offers many advantages. In particular, LSTMs maintain a tight control of what information is passed on from each timestep. This helps avoid vanishing or exploding gradients that often plague vanilla RNNs, allowing more effective training through backpropagation of long sequences. Previous work has even been done to show LSTMs ability to understand the underlying representation when trained on structured code written in C \cite{karpathy2015visualizing}. Further, LSTMs have been shown to be surprisingly effective at learning the structure behind both written language and raw C code. Karpathy et al. train an LSTM to predict the next character in a sample sequence of characters from the Linux source code, then look for interpretable activations within the LSTM. In doing so, they find LSTMs have the ability to understand the underlying structure of if statements, for loops, while loops, functions, and nesting among other things.

\begin{figure}[h]
\centering
\includegraphics[width=0.4\textwidth]{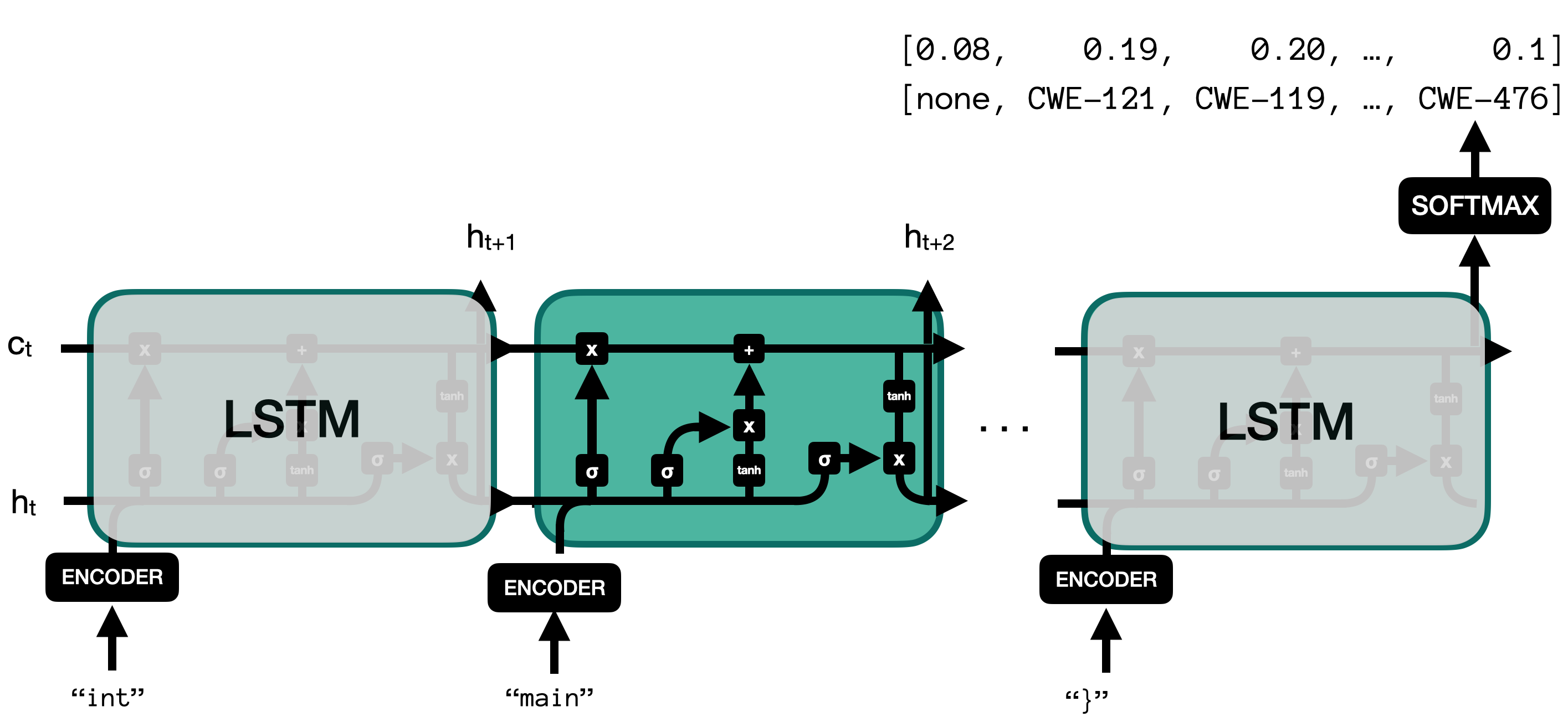}
\caption{Example of LSTM recurrence for predicting security vulnerabilities}\label{fig:lstm}
\end{figure}

Our LSTM vulnerability detection model is shown as in Fig. ~\ref{fig:lstm}. Each code symbol is an input to a LSTM cell has impact on its following symbols read in. The output vector of the last LSTM cell indicates the probabilities corresponding to each type of vulnerability contained in the input file.

\subsection{BiDirectional LSTM}
Traditional LSTMs going forward through a sequence, only see each input once and from one direction. This often means important information at the beginning of long sequences can be lost by the time the LSTM reaches the end of the sequence. BiDirectional LSTMs solve this problem by running two separate LSTMs, one that goes through the sequence from start to end and another that goes backward through the sequence from end to start \cite{bidirectionalLSTM}. After both LSTMs have finished going through the sequence, all the outputs of the forward and backward LSTMs are combined to make a final prediction, $\hat{y}$. 


For our purposes, BiDirectional LSTMs are particularly attractive because the sequences of text in the code files in our dataset tend to be relatively long compared to other common NLP tasks.

\subsection{BERT}
To sufficiently compare last NLP Transformer models with other traditional NLP deep learning models we use Bidirectional Encoder Representations from Transformers(BERT)\cite{devlin2018bert} as the representative due to its widespread popularity and bidirectional properties. Moreover, BERT has been shown to perform well at extracting information not only from English text but also languages with very different structure including Arabic \cite{antoun2020arabert} and Chinese \cite{cui2019pretraining}. These language and structure agnostic properties of BERT make it an attractive choice for applications outside of written language. BERT requires a fixed-length sequence as input, which poses a problem for the long sequences found in our security vulnerability dataset. We remedy this in a variety of ways which we describe in \ref{performance}.

The multi-head self-attention mechanism, which BERT makes heavy use of, is the key insight behind Transformers \cite{vaswani2017attention}. When each word is being processed, the self-attention mechanism looks at the other words in the sentence to see if they modify its meaning. For example, in the sentence "The dog drank water because it was thirsty," self-attention allows the word ``it" to be closely associated with the word ``dog" because ``it" is a reference to ``dog". 

\begin{figure}[h]
\centering
\includegraphics[width=0.4\textwidth]{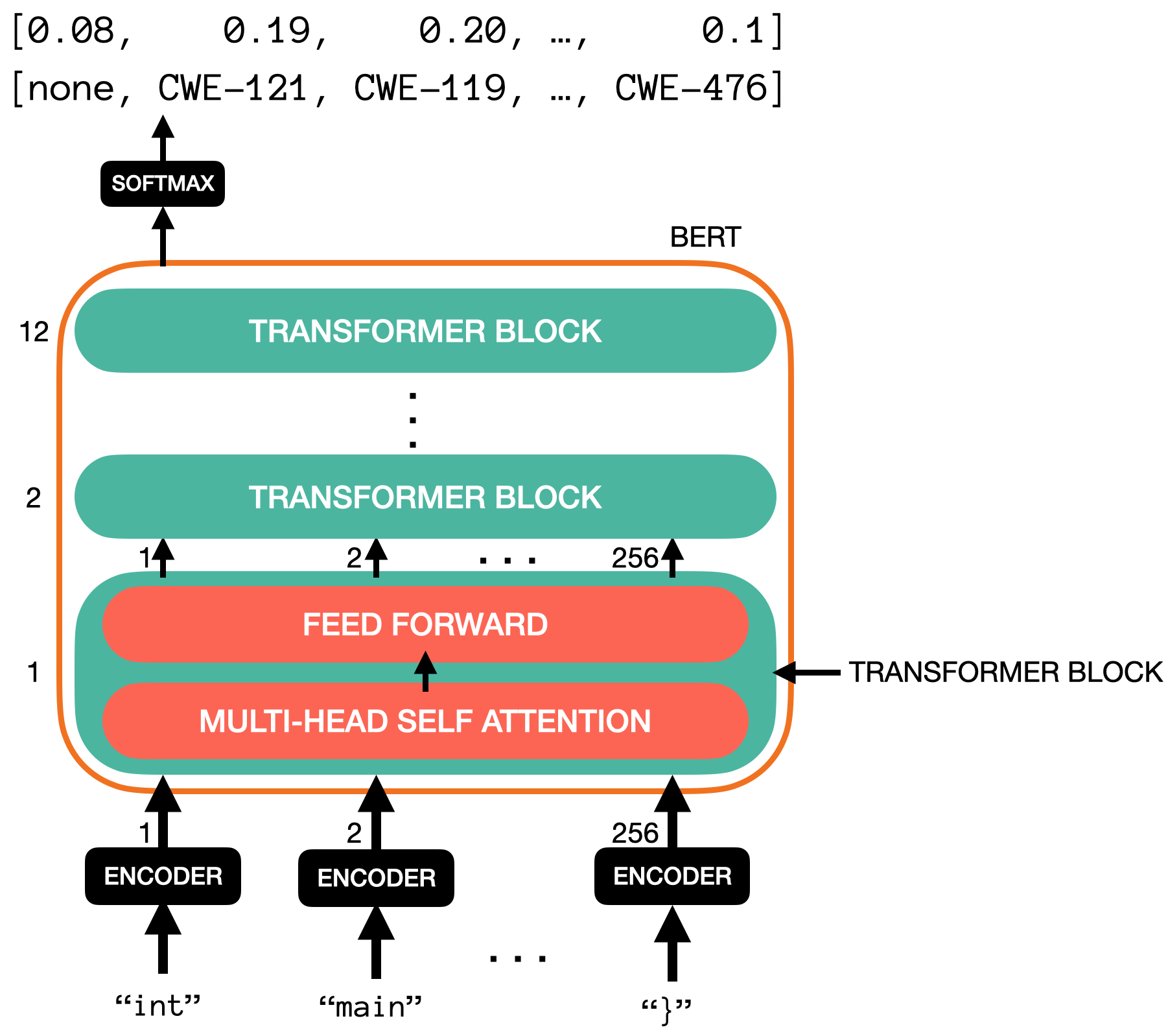}
\caption{BERT software vulnerability detection model with 12 Transformer blocks, each of which has 768 hidden units and 12 attention heads, with the input of 256 word vectors at a time.} \label{fig:bert}
\end{figure}

The architecture of BERT uses a series of Transformer blocks \cite{devlin2018bert}, which contain self-attention mechanisms, stacked on top of each other. Each Transformer block reads in a fixed length sentence and modifies or \textit{transforms} it into a new sentence of the same length. More specifically, it takes in a sentence, looks at the words of the sentence to modify their vector representations or meaning, and then feeds these series of modified vector representations through a regular feed-forward network for its final output. In our BERT software vulnerability detection model, there are 12 stacked Transformer blocks, each with a feed-forward network containing 768 hidden units and 12 attention heads. Our BERT model only takes in 256 word vectors at a time. These architecture parameters were chosen by \cite{devlin2018bert} to maintain consistent trainable model parameters with GPT\cite{Radford2018ImprovingLU} for architecture comparison purposes. The architecture diagram of our BERT detection model is shown in Fig. \ref{fig:bert}.

Along with self-attention, our BERT model uses a pre-training and fine-tuning process to take advantage of transfer learning in a similar way to \cite{howard2018universal}. During the pre-training phase, a BERT model is trained to predict a masked word from a given sentence in a dataset. In doing so, it learns to predict which words belong in a particular spot and in turn it learns the structure of the language. Due to the rigid syntax rules imposed by programming languages, we have observed that it is not necessary to pre-train our BERT model on computer code. Instead, our model was pre-trained on an English Wikipedia dataset containing 2.5 billion words which was used in \cite{devlin2018bert}, where it is known as BERT base.

After the pre-training phase, BERT is fine-tuned for a specific task. In our case, this task is text classification. BERT can only take in a fixed number of words at any given time. To do so, each file in our dataset is broken into smaller sequences where each sequence is made up of 256 words. At the end of the sequence, a classification score is given for each vulnerability class using the Softmax classifier and the vulnerability class with the highest score is effectively the vulnerability that the model has classified the file contains.

\section{Performance Evaluation}\label{performance}

\subsection{Experiment Settings}
Each experiment is run on a cluster of 8 Nvidia Volta Tesla V100s with 16 gigabytes of memory each, for two days. While the LSTM based models do not require two days to converge, the BERT based models continue to improve even beyond two days. Thus, to maintain a fair comparison between models, we run all models for the same duration on the same hardware.

All models are programmed in Python using the PyTorch library\cite{NEURIPS2019_9015}.

\subsection{Experiment Scenarios}
For our first experiment with traditional NLP methods, we use the unidirectional LSTM architecture. Due to only observing one token at a time, LSTMs allow for analyzing sequences of text of any arbitrary long length. This is unlike most transformer models, which can only observe a fixed length segment of tokens at any given time, which we talk about later in this section.

Our dataset contains text sequences that are relatively longer than most datasets, which can hinder the performance of LSTMs. To remedy this, we also experiment with BiDirectional LSTMs, which were expected to have modest performance improvements compared to unidirectional LSTMs by analysing the text sequence in both directions. 

For our first experiment with BERT, we use the technique similar to that which is described in \cite{alrfou2018characterlevel} where the full sequence is broken into fixed size segments which are then fed into the model. This is also known as the "vanilla" method in \cite{dai2019transformerxl}. For character prediction, this approach has the undesirable side effect where the model ignores the contextual information both before and after each segment. To remedy this, we add a vanilla RNN that takes in the output of each BERT segment. A softmax classifier is then applied to the very last RNN output for the final classification scores.

For our second experiment, we use a similar technique as before but instead feed the output of each segment into a unidirectional LSTM. With this approach, the contextual information is kept between each sequence at the LSTM level. Predictions are then made once after the last segment has gone through the BERT encoder and LSTM. The intuition behind this approach is that certain parts of code in the sequence refer to one another, and thus having context beyond each segment is useful for classifying security vulnerabilities.

For our last BERT experiment, we again use an LSTM but instead of unidirectional, we use a BiDirectional LSTM approach. With this approach, the model is allowed to evaluate the entire code sequence backwards as well as forwards, giving more context before a final classification is made.

Our results are summarized in the following Table \ref{tab:my_label}, with the most notable results highlighted in bold. The detail data analysis of these metrics are explained in the sections below. Note that although the RNN results are left out due to its performance being no better than random, we briefly describe RNN results in section \ref{sec:others}

\begin{table}[ht]
    \caption{Performance summary}
    \label{tab:my_label}
    \centering
    \begin{tabular}{|c|c|c|c|}
    \hline
     $\mathbf{Model}$ & $\mathbf{Accuracy}$ & $\mathbf{FLOPs}$ & $\mathbf{Unique Tokens}$\\
    \hline
    \hline
    LSTM & 72\% & 12.6M & 5.56m \\
    \hline
    BiDirectional LSTM & 79\% & 26M & 5.56m\\
    \hline
    BERT & 85\% & 133M & 32k\\
    \hline
    BERT + LSTM & $\mathbf{93.19\%}$ & 145M & 32k\\
    \hline
    BERT + BiLSTM & $\mathbf{93.49\%}$ & 159M & 32k\\
    \hline
    \end{tabular}
\end{table}

\subsection{Accuracy}
With respect to accuracy, the BERT model and its variants all outperformed the traditional NLP deep learning models that were tested. This is due to the attention mechanism in BERT models being able to extract context-aware information in a much better way than an LSTM or RNN alone. Further, the BERT+LSTM makes a large improvement on the vanilla BERT model, showing the importance of keeping contextual information across each segment. Somewhat surprisingly, by adding BiDirection to the LSTM+BERT gives only minor improvements in accuracy, while BiDirectional LSTM improves a unidirectional LSTM with the accuracy from 72\% to 79\%. Namely, the LSTM alone substantially improves with BiDirection, but the BERT+LSTM does not. This can be attributed to the fact that BERT already has a BiDirectional mechanism built in, and thus, adding another BiDirectional mechanism on top of it does not help the model extract any extra necessary information. 

The observation is on two folds. First, the introduction of bidirectional memory, which is either in BiDirectional LSTM or BERT, can significantly improve the accuracy performance of software vulnerability detection in the way of NLP because the contextual information is comprehensively considered. Second, the attention mechanism in BERT can result in another significant improvement on accuracy. With both mechanisms plus the LSTM, the accuracy can reach as high as over 93\%.

\subsection{FLOPs}
To compare the computational complexity of each model, we use Floating Point Operations(FLOPs) which allows comparison of computation without regard to hardware. The LSTM and BiDirectional LSTMs both use only 12.6M and 26M flops respectively whereas the transformer-based BERT models use an order of magnitude more FLOPs to evaluate the same text. Namely, the accuracy is at the cost of the computation. In practice, this means transformer-based models will take much more time to compute and will require more computational resources. Moreover, the relationship between FLOPs and accuracy indicates further work in more complex NLP architectures could further improve the accuracy of detecting security vulnerabilities.

\subsection{Unique Tokens}
The NLP deep learning models process the input text into tokens before processing further in their encoders. The tokenization algorithm, which splits long strings of text into separate words that can then be found in the encoder's dictionary, was the same for all of the traditional NLP deep learning models. If a model results in more tokens in its dictionary, more memory will be required to store these tokens during the computation. From the results, we can observe that because of the way these models break up longer strings into words, there are many more unique tokens with the traditional NLP deep learning models than with the BERT models, which were capped at 32k unique tokens. This indicates that the BERT models generalized much better than the traditional models, and, therefore, expect to perform better when subjected to data outside our dataset.



\section{Conclusions}
\label{sec:others}
In this work, we have first developed a novel dataset of preprocessed software vulnerabilities. Then, we have designed a few deep learning models to detect and classify software vulnerabilities in source code in its text presentation based on recent NLP deep learning research advance. With extensive experiments, it is shown that Transformers NLP deep learning models can achieve impressive results in detecting and classifying software vulnerabilities in code. Moreover, we observe that it is very important to maintain the contextual information across the entire sequence for this particular task. The built dataset of software vulnerabilities can also be used by other researchers to further improve this field

\section{Acknowledgment}
This material is based upon work supported by the National Science Foundation under Grant No. \#1923712 and a grant by Cisco through the NSF I/UCRC S2ERC research center. 

\bibliographystyle{unsrt}  
\bibliography{references}  

\end{document}